\pdfoutput=1
\documentclass[10pt,twocolumn]{article} 
\usepackage{webstructurestyle}
\usepackage{times}
\usepackage{graphicx}
\usepackage{epstopdf}
\usepackage{amssymb}
\usepackage{url,hyperref}
\usepackage{subcaption}
\usepackage{makecell}
\usepackage{booktabs,multirow}
\usepackage{array}

\begin{document}

\title{Analysis of the Web Graph Aggregated by Host and Pay-Level Domain }

\author{Agostino Funel \\
ENEA, Italy \\
\\
agostino.funel@enea.it  \\
}

\maketitle
\thispagestyle{empty}

\begin{abstract}
In this paper the web is analyzed as a graph aggregated by host and pay-level domain (PLD). The web graph datasets, publicly available, have been released by the Common Crawl Foundation~\footnote{http://commoncrawl.org} and are based on a web crawl performed during the period May-June-July 2017. The host graph has $\sim$1.3 billion nodes and $\sim$5.3 billion arcs. The PLD graph has $\sim$91 million nodes and $\sim$1.1 billion arcs. We study the distributions of degree and sizes of strongly/weakly connected components (SCC/WCC) focusing on power laws detection using statistical methods. The statistical plausibility of the power law model is compared with that of several alternative distributions.  While there is no evidence of power law tails on host level, they emerge on PLD aggregation for indegree, SCC and WCC size distributions. Finally, we analyze distance-related features by studying the cumulative distributions of the shortest path lengths, and give an estimation of the diameters of the graphs. 
\end{abstract}

\section{Introduction}
The web is a complex system and its graph structure can be analyzed on different levels of granularity. 

 The page level is the World Wide Web (WWW) where a node is a webpage uniquely identified by its URL  and an arc is a hypertextual link between two webpages. Many webpages can be hosted on a computer (web server) uniquely identified by an IP address. On the host level each node of the web graph is a web server and two nodes are connected by an arc if exists at least a hypertextual link between the webpages hosted on the corresponding servers. Similarly, a next level of aggregation can be obtained by grouping web servers into Internet domains. In this case a node is a domain  and there is an arc between two domains if exist at least two servers, each belonging to one of the domains, connected by an arc. The greater the level of aggregation the greater the scale at which the web is observed. A more precise definition of these levels will be given in Section~\ref{sez_datasets}. Knowledge of the web structure is not only fascinating in itself but also important for many reasons. For example, it can help search engine developers to find better ranking algorithms; it can be useful to design efficient crawling strategies and to predict the connectivity of the web in the case of a widespread disconnection of nodes or break of links.

The main purpose of this work is to provide additional information on the web graph structure. In particular, we show that there is no statistical support  to affirm that on host level the heavy-tailed distributions of degree and sizes of the components of the web graph are power laws. Power laws emerge only on the macroscopic scale of PLD aggregation. It is an unanswered question which is the real mathematical form of the distributions of degree and components in the web. In this work we compare the power law  with several alternative  models by using statistical methods. Finally we analyze the connectivity of the web graph on host and PLD aggregation by studying the cumulative distributions of the shortest path lengths, and give an estimation of the diameters of the graphs.

The paper is organized as follows: in Section~\ref{sez_relwork} we present the results of previous analysis of the web graph structure; in Section~\ref{sez_datasets} we describe the datasets used in this work and introduce the terms which define the levels of aggregation; in Section~\ref{sez_method} we explain the methodology of analysis.  Sections~\ref{sez_analysis_hl} and~\ref{sez_analysis_dl} report the analysis of the web graph on host and PLD aggregation, respectively. In Section~\ref{sez_conclusion} we summarize the main results and make the final observations.

\section{Related Work}\label{sez_relwork}
The web graph structure has been studied by many authors.  
In the work of Broder {\itshape et al}.~\cite{Broder2000} the web graph structure was analyzed on the page level using two Altavista crawls (performed in May and October 1999) each with $\sim$200 million pages and $\sim$1.5 billion links. They found that the in and out degree distributions are power law with exponent $\sim$2.10 and $\sim$2.72 respectively, in agreement (in the case of indegree) with the theoretical predictions of Kumar {\itshape et al}.~\cite{Kumar99_1} and Barabasi and Albert~\cite{Barabasi_Albert_99}. At macroscopic scale they found a single WCC containing over 90\% of the nodes. This WCC breaks in four pieces (roughly of the same size) forming a "bow-tie" structure at the heart of which there is a giant SCC which contains $\sim$28\% of the nodes. The other pieces are called IN, OUT and TENDRILS. IN contains pages that can reach the SCC but can not be reached from it. OUT contains pages that can be reached from SCC but do not link back to it. TENDRILS contains pages completely disconnected from the SCC. The diameter of the central SCC was estimated to be at least 28 and that of the whole graph at least 500 and likely to be over 900. The percentage of connected pairs of nodes and their average distance were estimated $\sim$25\% and $\sim$16 respectively. They also found a power law distribution with exponent $\sim$2.54 for the sizes of the SCC and WCC, claiming it is a basic web property. However, no details were reported about the statistical plausibility of the power laws.
\\ \indent The web is a very large complex system and a detailed knowledge of its structure could be obtained only by crawling it completely. Moreover, the crawling process itself affects the global picture of the web~\cite{Serrano2007},~\cite{Donato2005},~\cite{Zhu2008}. 
\\ \indent A larger dataset was analyzed by Meusel {\itshape et al}.~\cite{Meusel2015} from a crawl, provided by the Common Crawl Foundation, gathered in the first half of 2012 and released in August of the same year. Their analysis was conducted on page, host and PLD levels. The sizes (number of nodes/arcs in billion) of the graphs being respectively $\sim$ (3.6/128.7), $\sim$(0.1/2.0), $\sim$(0.04/0.6). They used statistical methods to test power laws and proved that the distributions of in/out degree and sizes of SCC/WCC are not power laws on page and host level, while power laws emerge for the indegree and PageRank distributions  of the PLD graph. This would suggest that power laws might be due to aggregation or crawling artifacts rather than being a structural property of the web,  living open the question which is the correct mathematical form of the distributions. Authors estimated also the percentage of pairs of connected nodes, their average distance and the lower bound of the diameters. The page graph has $\sim$48\% of connected pairs with average distance $\sim$13, resulting more connected than what estimated by Broder {\itshape et al}, and a diameter of at least 5282. The host graph has $\sim$36\% of connected pairs with average distance $\sim$5, and a diameter at least 261. The PLD graph has $\sim$42\% of connected pairs with average distance $\sim$4 and a diameter at least 48.
\\ \indent Inspired by this work, we perform the same statistical goodness of fit analysis on larger host and PLD graphs and, in addition, we examine the statistical plausibility of other models of distributions alternative to the power law.

\section{Datasets and definitions}\label{sez_datasets}
The host and PLD web graphs, publicly available, are provided by the Common Crawl Foundation. These graphs have been extracted from a web crawl which gathered data during the period May-June-July 2017. In Table~\ref{tab_wgsizes} are shown the sizes of the datasets.

\begin{table}[htbp]
\begin{center}
\begin{tabular}{|c|c|c|}
\hline
Aggregation & \# Nodes  & \# Arcs  \\
\hline
\hline
Host     & 1306661614  & 5268397861 \\
\hline
PLD     & 91034128 &   1071173924 \\
\hline
\end{tabular}
\end{center}
\caption{Host and PLD web graph sizes.}
\label{tab_wgsizes}
\end{table}

\begin{description}
  \item[Host:] The name or address of a web server can be extracted by its URL by excluding protocol, authentication, port, path, query and fragment substrings. For example the web servers of the URLs http://www.example.1.com and http://www.foo.a.b.co.uk:8080/path?query=answer are  www.example.1.com and www.foo.a.b.co.uk.
  \item[PLD:] This level of aggregation is based on the Public Suffix List, an initiative of Mozilla.~\footnote{https://www.publicsuffix.org/} It is a catalog of Internet domain name suffixes that can be directly registered by users. The PLD of a host is obtained by aggregating one dot above the  public suffix. The PLD of the hosts of the example above are 1.com and b.co.uk because .com and .co.uk are on the Public Suffix List.   
\end{description}

The host graph includes $\sim$1.2 billion nodes that have been identified as targets of a link from  a crawled page. The number of domain name registrations at the end of the second quarter of 2017 was $\sim$331.9 millions~\footnote{https://www.verisign.com/assets/domain-name-report-Q22017.pdf} thus our PLD dataset contains $\sim$27.4\% of all PLDs registered at that time.

\section{Methodology of analysis}\label{sez_method}
For host and PLD aggregation we study the distributions of in/out degree and sizes of SCC and WCC and for each of them make a best fit to a power law. In order to decide the statistical plausibility of a power law we follow the procedure described in the work of Clauset {\itshape et al.}~\cite{Clauset2009}. The best fit power law parameters $x_{min}$ and $\alpha$ are calculated with the method of maximum likelihood. After that a goodness of fit test based on the Kolmogorov-Smirnov statistic provides a  $p$-value. If $0 \leq p < 0.1$ the power law hypothesis is rejected if $0.1 \leq p \leq 1$ it is accepted. In the case of a power law detection as further check we compare the experimental data with synthetic data randomly generated from a power law with the same parameters of the detected one. Of course, there could be other distributions which might fit better the data. We test, both in the discrete and continuous fit formalism, the power law ($pl$) hypothesis against competing models and choose: exponential ($exp$), lognormal ($logn$), truncated power law ($tpl$) and stretched exponential ($sexp$). The degree and the size of a graph component are integers and the discrete formalism is more appropriate for studying the related distributions. However, for a very large graph the number of elements of its distributions is huge and the continuous formalism, which is computationally less intensive, provides a good approximation and might give additional clues on the properties of the data distributions.   The continuous form of the tested models is shown in Table~\ref{tab_models}.

\begin{table}[ht]
\begin{center}
\begin{tabular}{c|c}
power law     & $x^{-\alpha}$   \\
\hline
\hline
alternative model & $f(x)$  \\
\hline
exponential    & $e^{-\lambda x}$  \\
lognormal  & $\frac{1}{x}exp\left[-\frac{\left( ln x - \mu \right)^2}{2\sigma^2}\right]$ \\
truncated power law & $x^{-\alpha}e^{-\lambda x}$ \\
stretched exponential & $x^{\beta - 1}e^{-\lambda x^{\beta}}$ \\
\end{tabular}
\end{center}
\caption{Models of distributions $p(x)=C f(x)$ whose statistical plausibility is compared. The constant $C$ is obtained by the normalization $\int_{x_{min}}^{\infty} C f(x)=1$.}
\label{tab_models}
\end{table}

The comparison between two distributions $f_A$ and $f_B$ is achieved by calculating the (normalized) loglikelihood ratio $R(f_A/f_B)$. The sign of $R$ decides which is the best model: if $R > 0$ ($R < 0$) then $f_A$ ($f_B$) is the favorite distribution. If $R = 0$ there is not a favorite model with respect to the other. The statistical significance of the sign of $R$ depends on a  $q$-value. If $0 \leq q < 0.1$ the sign of $R$ is a reliable indicator of which model is the best one. If $0.1 \leq q \leq 1$ neither of the two models is favorite.  Based on these observations we can judge the statistical plausibility of $f$ as a better fit distribution compared to a power law as shown in Table~\ref{tab_llr}. If there are many potential candidates we compare them pairwise to find out which one should be considered the best alternative to the power law. The software we use in this work are: {\ttfamily plfit}~\footnote{http://github.com/ntamas/plfit} for the goodness of fit tests; {\ttfamily powerlaw}~\footnote{https://pypi.python.org/pypi/powerlaw}, described in the work of Alstott {\itshape et al}.~\cite{Alstott2014}, for the comparison of models and the {\ttfamily SNAP} library~\cite{leskovec2016snap} for the analysis of the structural properties of the graphs.

\begin{table}[ht]
\begin{center}
\begin{tabular}{|c|c|c|}
\hline
    $R(pl/f) > 0$ & $0 \leq q < 0.1$ & none\\
\hline
$-\infty < R(pl/f) < -\infty$ & $0.1 \leq q \leq 1$ & undecidable\\
\hline
$R(pl/f) = 0$ & $0 \leq q \leq 1$ & undecidable\\
\hline
$R(pl/f) < 0$ & $0 \leq q < 0.1$ & strong\\
    \hline
\end{tabular}
\end{center}
\caption{Statistical plausibility of an alternative distribution $f$ compared with a power law based on the result of the likelihood ratio test.} 
\label{tab_llr}
\end{table}

We also study the cumulative distributions of the shortest path lengths which are very computational intensive.  The {\ttfamily SNAP} library adopts a fast and memory-efficient algorithm based on an approximation of the neighbourhood function as described in the work of  Palmer {\itshape et al}.~\cite{Palmer2002}. In order to estimate the lower bound of the diameters we perform a breadth first search (BFS) over the graphs  using 10000 starting test nodes.

\section{Analysis of the host graph}~\label{sez_analysis_hl}
The host graph has $\sim$1.3 billion nodes and $\sim$5.3 billion arcs. There are 16903 zero degree nodes ($\sim$0.0013\% of the total). The average degree is $\sim$8.06.

\subsection{Degree distributions}
\begin{figure}[!h]
  \begin{center}
    \includegraphics[width=3.5in]{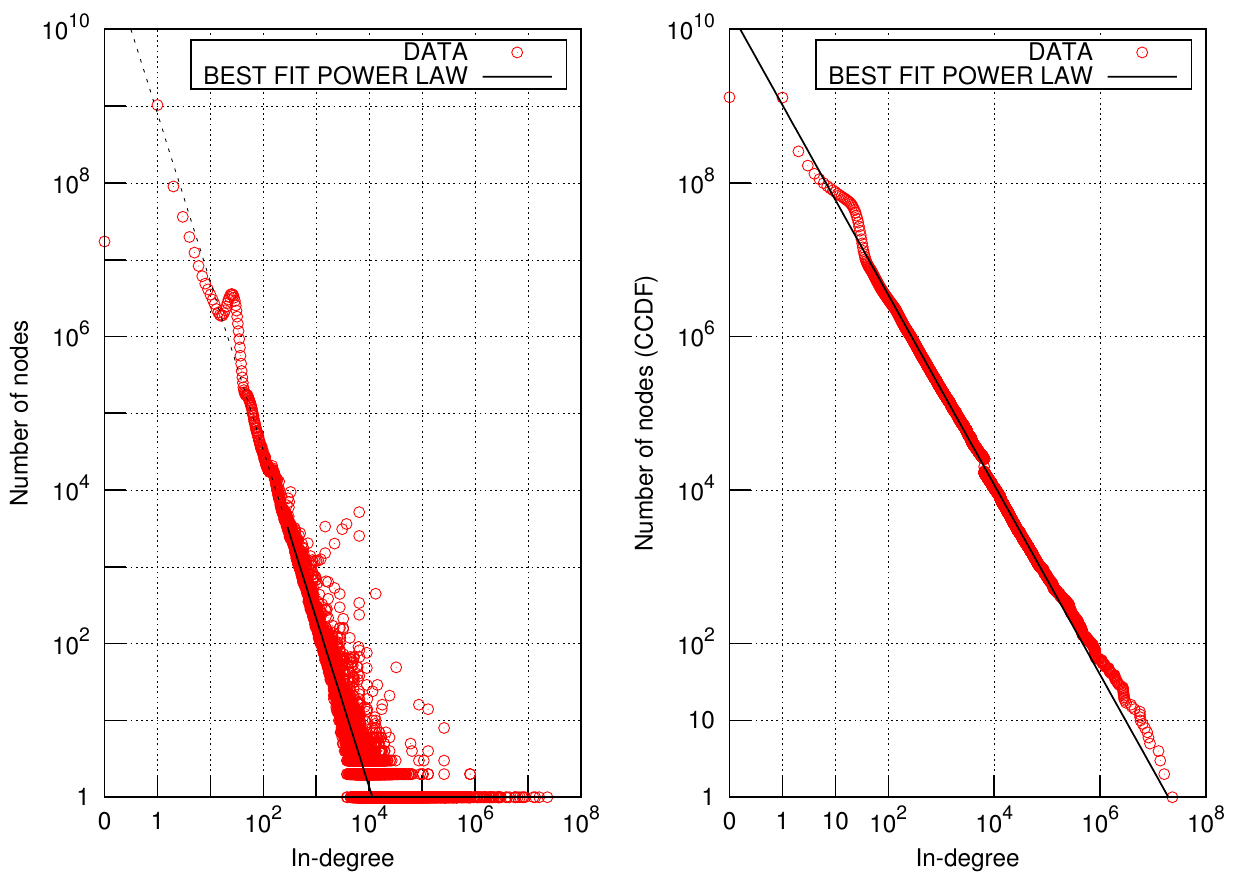}
\caption{Host graph indegree distribution (left) and its CCDF (right). The best fit power law parameters are: $x_{min}=250$, $\alpha=2.193\pm0.001$. The $p$-value is $0.00\pm 0.01$. }
\label{fig_in_ccd_deg_hl2017050607}
  \end{center}
\end{figure}

In Figure~\ref{fig_in_ccd_deg_hl2017050607} are shown the frequency plots of the indegree distribution and its complementary cumulative density function (CCDF) in log-log scale. In order to plot also the value of the distributions for the nodes with zero degree we manually shift the point $x = 0$ to 0.1 and label it 0 on the X-axis.

The indegree distribution has a hump centered at $x\simeq25$ with $\sim$4 million nodes. This peak distorts the initial linear shape and is clearly visible as a concavity of the CCDF. In the region $10^3 \leq x \leq 10^6$ there are spikes which might indicate a deviation from power law. In fact, despite the shape of the CCDF appears almost linear, from the goodness of fit test we obtain $p=0.00\pm 0.01$. The best fit power law parameters are $x_{min}=250$ and $\alpha=2.193\pm0.001$. The nodes in the region $x \geq x_{min}$ are $\sim$0.082\% of the distribution. The dashed (solid) part of the black line in a plot of a distribution  indicates the region $x < x_{min}$ ($x \geq x_{min}$). The number of nodes with indegree equal to zero is $\simeq 1.73 \cdot 10^7$ ($\sim$1.32\% of the distribution). The maximum indegree is 23055296.

In Table~\ref{tab_compare_discrete_cont_indeghl} are shown the results of the comparison between the power law and alternative models.  In the case of the discrete formalism  we find strong support for the lognormal while if we use the continuous form of the distributions for the fit we find that none of the tested models can be considered a plausible alternative to the power law. 

\begin{table}[htbp]
\begin{center}
\resizebox{\columnwidth}{!}{ 
\begin{tabular}{c|c|c|c}
\multicolumn{4}{c}{discrete fit indegree host graph}\\
\hline
\hline
 \boldmath{$f$} & \boldmath{$R(pl/f)$} & \boldmath{$q$} & \makecell{\textbf{statistical plausibility of} \boldmath{$f$}\\ \textbf{as alternative to the power law}} \\
\hline
$exp$	&	74.652299 &	0 &	none \\
\hline
$logn$ &	-1.8227830 &	0.067575 &	strong \\
\hline
$tpl$ & 	-0.584161 &	0.512644 &	undecidable \\
\hline
$sexp$ &	15.742337 &	0 &	none \\
\hline
\hline
\boldmath{$f_A/f_B$} &	\boldmath{$R(f_A/f_B)$} & \boldmath{$q$} & \textbf{comment} \\
\hline
$logn/tpl$ &	1.829324 &	0.067351 &	strong support for the $logn$  \\ 
\end{tabular}
}
\end{center}
\begin{center}
\resizebox{\columnwidth}{!}{
\begin{tabular}{c|c|c|c}
\multicolumn{4}{c}{continuous fit indegree host graph}\\
\hline
\hline
 \boldmath{$f$} & \boldmath{$R(pl/f)$} & \boldmath{$q$} & \makecell{\textbf{statistical plausibility of} \boldmath{$f$}\\ \textbf{as alternative to the power law}} \\
\hline
$exp$   &       24.766065 &     0 &     none \\
\hline
$logn$ &        0.597332 &    0.550286 &      undecidable \\
\hline
$tpl$ &         -0.516431 &     0.629009 &      undecidable \\
\hline
$sexp$ &        43.152895 &     0 &     none \\
\hline
\hline
\boldmath{$f_A/f_B$} &  \boldmath{$R(f_A/f_B)$} & \boldmath{$q$} & \textbf{comment} \\
\hline
$logn/tpl$ &    -0.680971 &      0.495890 &      none of the tested models is favorite  \\
\end{tabular}
}
\end{center}
\caption{Results of the likelihood ratio test for the indegree distribution of the host graph.  Discrete fit: the lognormal is the most statistically plausible alternative to the power law among all tested models. Continuous fit: none of the tested models is statistically plausible as alternative to the power law. }
\label{tab_compare_discrete_cont_indeghl}
\end{table}

\begin{figure}[htbp]
  \begin{center}
    \includegraphics[width=3.5in]{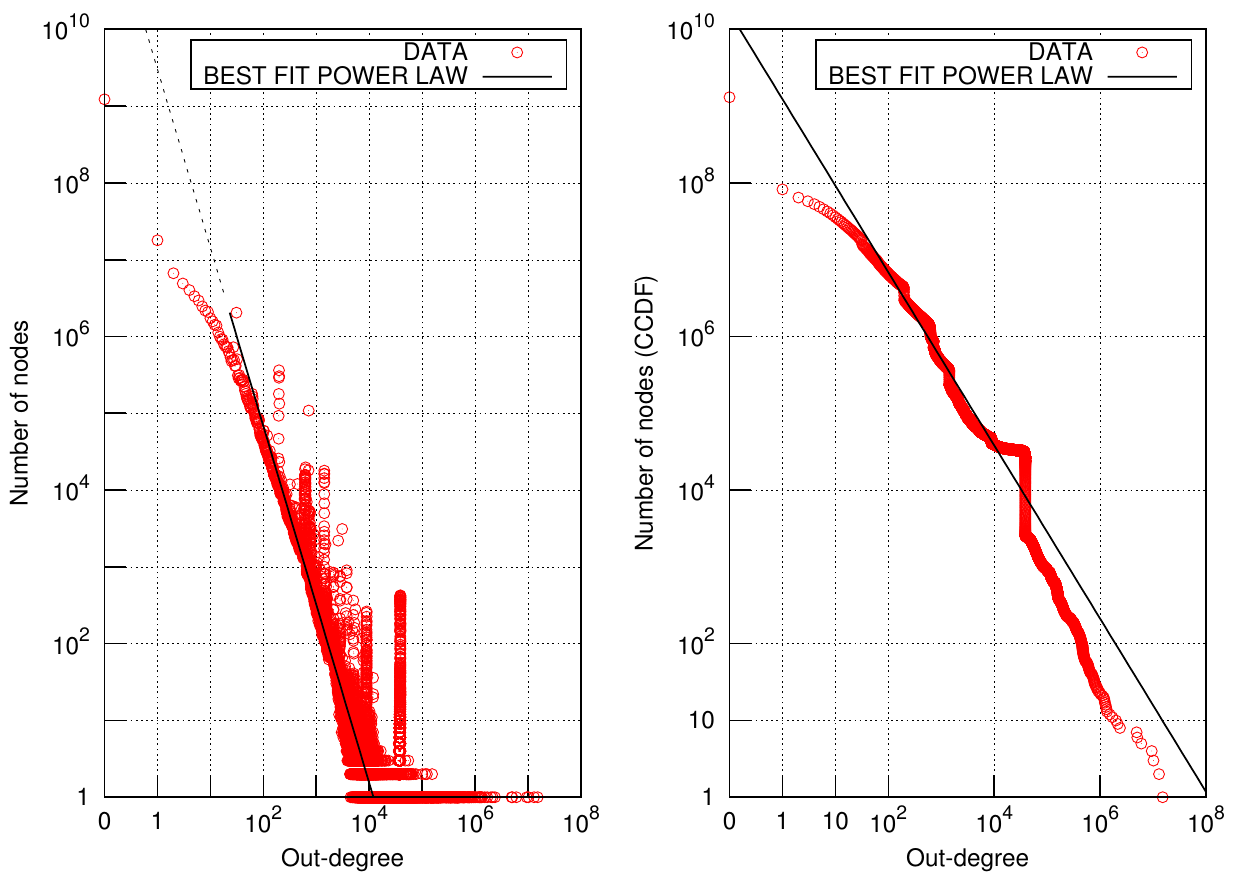}
\caption{Host graph outdegree distribution (left) and its CCDF (right). The best fit power law parameters are: $x_{min}=23$, $\alpha=2.3242\pm0.0001$. The $p$-value is $0.00\pm 0.01$. }
\label{fig_out_ccd_deg_hl2017050607}
  \end{center}
\end{figure}

In Figure~\ref{fig_out_ccd_deg_hl2017050607} are shown the frequency plots of the outdegree distribution and its CCDF.  
The nodes with outdegree equal to zero are $\sim$93\% of the distribution and their number is $\sim$1.2$\cdot 10^9$ which is the number of dangling hosts included during the crawling process. These nodes are not directly gathered by the crawler yet are pointed to from a link on a crawled page.  In the region $1 \leq x \leq 30$ there is a concavity. There are many spikes in the region $200 \leq x \leq 40000$ which however contains $\sim$0.24\% of the distribution. The notched shape of the outdegree distribution causes the shape of the CCDF to be highly non linear. The best fit power law parameters are $x_{min} = 23$ and $\alpha = 2.3242 \pm 0.0001$. The region $x \geq x_{min}$ contains $\sim$1.69\% of the distribution. The $p$-value is $0.00\pm 0.01$. The maximum outdegree is 15090917.

\begin{table}[!h]
\begin{center}
\resizebox{\columnwidth}{!}{ 
\begin{tabular}{c|c|c|c}
\multicolumn{4}{c}{discrete fit outdegree host graph}\\
\hline
\hline
 \boldmath{$f$} & \boldmath{$R(pl/f)$} & \boldmath{$q$} & \makecell{\textbf{statistical plausibility of} \boldmath{$f$}\\ \textbf{as alternative to the power law}} \\
\hline
$exp$   &       455.329430 &     0 &     none \\
\hline
$logn$ &        -239.275758 &    0 &      strong \\
\hline
$tpl$ &         -36.993154 &     0 &      strong \\
\hline
$sexp$ &        3.288489 &     0.001 &     none \\
\hline
\hline
\boldmath{$f_A/f_B$} &  \boldmath{$R(f_A/f_B)$} & \boldmath{$q$} & \textbf{comment} \\
\hline
$logn/tpl$ &    52379.86440 &      0 &      strong support for the $logn$  \\
\end{tabular}
}
\end{center}
\begin{center}
\resizebox{\columnwidth}{!}{
\begin{tabular}{c|c|c|c}
\multicolumn{4}{c}{continuous fit outdegree host graph}\\
\hline
\hline
 \boldmath{$f$} & \boldmath{$R(pl/f)$} & \boldmath{$q$} & \makecell{\textbf{statistical plausibility of} \boldmath{$f$}\\ \textbf{as alternative to the power law}} \\
\hline
$exp$   &       141.351591 &     0 &     none \\
\hline
$logn$ &        -187.661597 &    0 &      strong \\
\hline
$tpl$ &         -34.653582 &     0 &      strong \\
\hline
$sexp$ &        -87.928342 &     0 &     strong \\
\hline
\hline
\boldmath{$f_A/f_B$} &  \boldmath{$R(f_A/f_B)$} & \boldmath{$q$} & \textbf{comment} \\
\hline
$tpl/sexp$ &    -14.999285 &    0 &     $sexp$ most plausible than $tpl$ \\
\hline
$logn/sexp$ &    48.827806 &      0 &      strong support for the $logn$ \\
\end{tabular}
}
\end{center}
\caption{Results of the likelihood ratio test for the outdegree distribution of the host graph. Both the discrete and continuous fit calculations assign to the lognormal the strongest support.}
\label{tab_compare_discrete_cont_outdeghl}
\end{table}

The results of the comparison between the power law and other models are shown in Table~\ref{tab_compare_discrete_cont_outdeghl}. In the discrete fit calculations the lognormal and the truncated power law are the eligible candidates  however from their comparison the lognormal is the most plausible. In the continuous fit calculations there are three models, lognormal, truncated power law and stretched exponential as eligible alternative to the power law. However from their comparison it results that the lognormal is the best alternative. It is interesting to note that in both cases the lognormal has the strongest support.

\subsection{Components}
\begin{figure}[htbp]
  \begin{center}
    \includegraphics[width=3.5in]{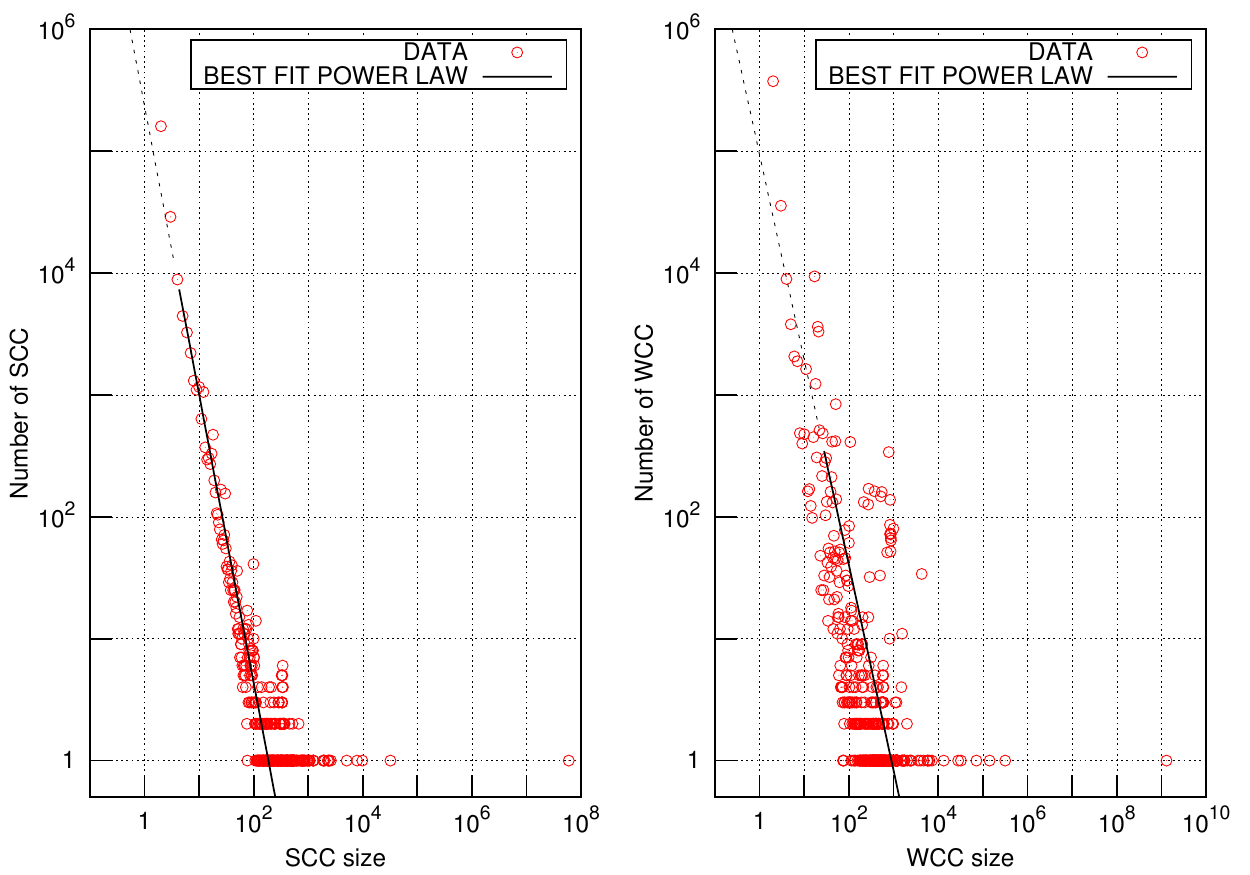}
\caption{SCC (left) and WCC (right) size distributions of the host graph. The best fit power law parameters for the SCC are: $x_{min}=4$, $\alpha = 2.367\pm0.005$. For the WCC we have $x_{min}=22$, $\alpha = 1.684\pm0.001$. The two distributions have $p=0.00\pm0.01$.}
\label{fig_scc_wcc_hl2017050607}
  \end{center}
\end{figure}

Almost all nodes of the host graph are weakly connected. The fraction of nodes in the largest WCC is $\sim$99.7\%. The largest SCC is considerably smaller and contains $\sim$4.5\% of the nodes.   In the analysis of Meusel {\itshape et al}.  the largest WCC and SCC contain respectively $\sim$87\% and $\sim$47\% of the whole host graph whose size, however, is about one order of magnitude smaller than that of the host graph analyzed in this work. The marked difference between the two sizes of the largest SCC could be due to different methodologies in the process of graph extraction from the gathered data. However, this analysis confirms the presence of a giant WCC in the host graph.          
In Figure~\ref{fig_scc_wcc_hl2017050607} are shown the distributions of the sizes of the SCC and WCC. 

The best fit power law parameters  of the SCC distribution are $x_{min}=4$, $\alpha = 2.367\pm0.005$. The $p$-values is $0.00\pm0.01$. The region $x \geq x_{min}$ covers $\sim$13.2\% of the distribution.

For the WCC distribution we have $x_{min}=22$, $\alpha = 1.684\pm0.001$ and $p=0.00\pm0.01$. The region $x \geq x_{min}$ covers $\sim$2\% of the distribution.

A visual inspection shows that the points of the WCC distributions are widely spreaded around the best fit power law line while the ones of the SCC are not. However from the goodness of fit test we ascertain that  for neither of the two distributions the power law model is statistically plausible.
 
Also for the SCC and WCC distributions we test alternative models to the power law. The results are shown in Table~\ref{tab_compare_discrete_cont_scchl} and~\ref{tab_compare_discrete_cont_wcchl} . While from the continuous fit calculations we infer that none of the tested models has statistical support, the calculations in the discrete formalism indicate that both for the SCC and WCC distributions the lognormal is the best alternative.

\begin{table}[htbp]
\begin{center}
\resizebox{\columnwidth}{!}{ 
\begin{tabular}{c|c|c|c}
\multicolumn{4}{c}{discrete fit SCC host graph}\\
\hline
\hline
 \boldmath{$f$} & \boldmath{$R(pl/f)$} & \boldmath{$q$} & \makecell{\textbf{statistical plausibility of} \boldmath{$f$}\\ \textbf{as alternative to the power law}} \\
\hline
$exp$   &       13.295007 &     0 &     none \\
\hline
$logn$ &        -2.677685 &    0.007413 &      strong \\
\hline
$tpl$ &         -1.323552 &     0.005272 &      strong \\
\hline
$sexp$ &        7.573920 &     0.001 &     none \\
\hline
\hline
\boldmath{$f_A/f_B$} &  \boldmath{$R(f_A/f_B)$} & \boldmath{$q$} & \textbf{comment} \\
\hline
$logn/tpl$ &    6.068555 &      0 &      strong support for the $logn$  \\
\end{tabular}
}
\end{center}
\begin{center}
\resizebox{\columnwidth}{!}{ 
\begin{tabular}{c|c|c|c}
\multicolumn{4}{c}{continuous fit SCC host graph}\\
\hline
\hline
 \boldmath{$f$} & \boldmath{$R(pl/f)$} & \boldmath{$q$} & \makecell{\textbf{statistical plausibility of} \boldmath{$f$}\\ \textbf{as alternative to the power law}} \\
\hline
$exp$   &       6.281507 &     0 &     none \\
\hline
$logn$ &        1.85880 &    0.063044 &      none \\
\hline
$tpl$ &         1.110213 &     0.009236 &      none \\
\hline
$sexp$ &        7.328739 &     0 &     none \\
\hline
\end{tabular}
}
\end{center}
\caption{Results of the likelihood ratio test for the size distribution of the SCC of the host graph.  Discrete fit: the lognormal is the most statistically plausible alternative to the power law among all tested models. Continuous fit: none of the tested  models can be considered  a statistically plausible alternative to the power law. }
\label{tab_compare_discrete_cont_scchl}
\end{table}

\begin{table}[!h]
\begin{center}
\resizebox{\columnwidth}{!}{ 
\begin{tabular}{c|c|c|c}
\multicolumn{4}{c}{discrete fit WCC host graph}\\
\hline
\hline
 \boldmath{$f$} & \boldmath{$R(pl/f)$} & \boldmath{$q$} & \makecell{\textbf{statistical plausibility of} \boldmath{$f$}\\ \textbf{as alternative to the power law}} \\
\hline
$exp$   &       4.933097 &     0 &     none \\
\hline
$logn$ &        -7.670585 &    0 &      strong \\
\hline
$tpl$ &         0.713476 &     0 &      none \\
\hline
$sexp$ &        103.577509 &     0 &     none \\
\hline
\end{tabular}
}
\end{center}
\begin{center}
\resizebox{\columnwidth}{!}{ 
\begin{tabular}{c|c|c|c}
\multicolumn{4}{c}{continuous fit WCC host graph}\\
\hline
\hline
 \boldmath{$f$} & \boldmath{$R(pl/f)$} & \boldmath{$q$} & \makecell{\textbf{statistical plausibility of} \boldmath{$f$}\\ \textbf{as alternative to the power law}} \\
\hline
$exp$   &       8.466598 &     0 &     none \\
\hline
$logn$ &        182.567972 &    0 &      none \\
\hline
$tpl$ &         0.007864 &     0.962260 &       undecidable \\
\hline
$sexp$ &        160.308865 &     0 &     none \\
\hline
\end{tabular}
}
\end{center}
\caption{Results of the likelihood ratio test for the size distribution of the WCC of the host graph. Discrete fit: the lognormal is the most statistically plausible alternative to the power law among all tested models. Continuous fit: none of the tested  models can be considered a statistically plausible alternative to the power law. }
\label{tab_compare_discrete_cont_wcchl}
\end{table}

\subsection{Distances and diameters}
\begin{figure}[htbp]
  \begin{center}
    \includegraphics[width=3.5in]{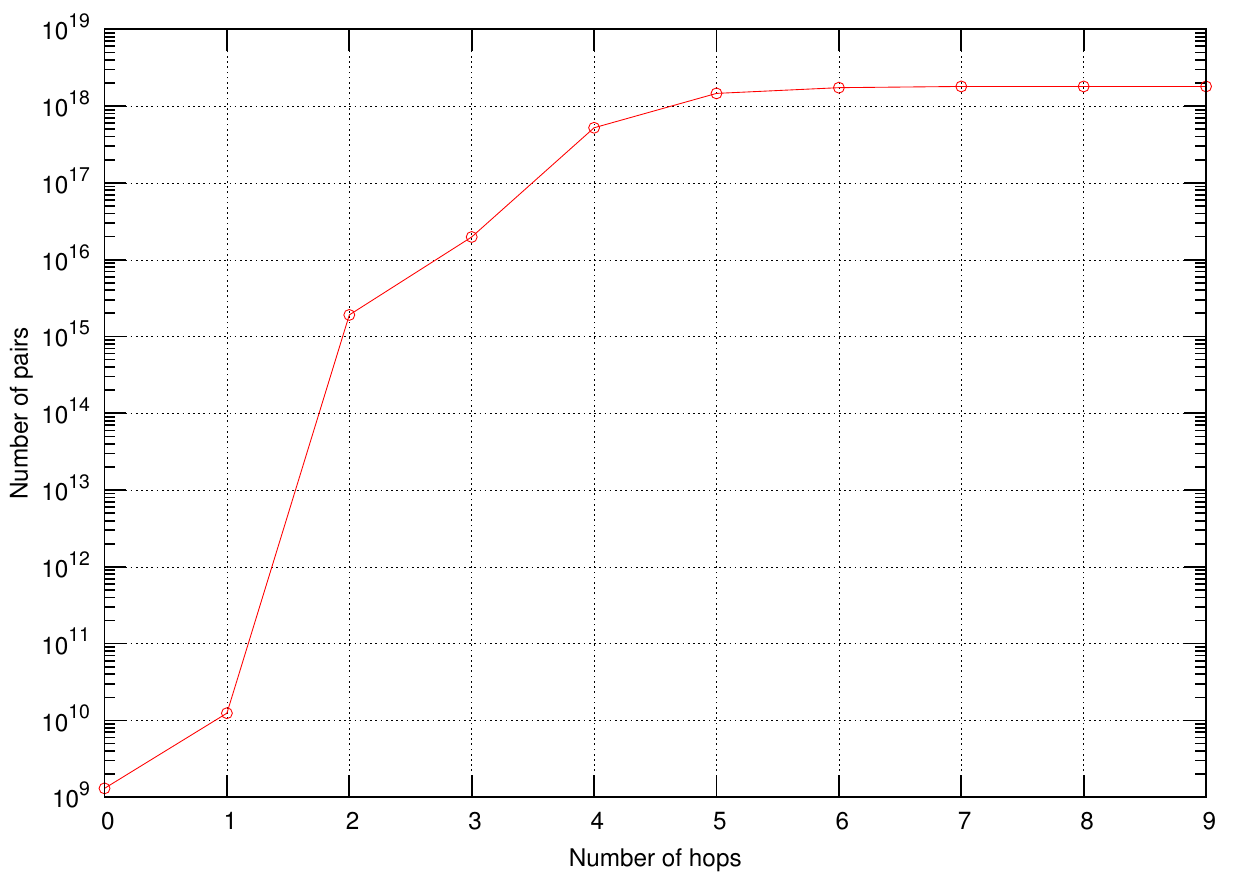}
\caption{Cumulative distribution of the shortest path lengths of the host graph. The Y-axis shows $N(h)$, the number of pairs of nodes with distance within $h$ hops. The effective diameter is $5.6\pm0.6$.}
\label{fig_diam_hl2017050607}
  \end{center}
\end{figure}

In the host graph $\sim$90\% of all pairs of nodes have distance within $5.6\pm0.6$ as shown in Figure~\ref{fig_diam_hl2017050607} where is plotted the cumulative distribution of the shortest path lengths (hop plot). The lower bound of the full diameter, estimated with a BFS algorithm with 10000 random starting nodes, is 970.

\section{Analysis of the PLD graph}~\label{sez_analysis_dl}
The PLD graph has $\sim$9.1 millions nodes and $\sim$1.1 billion arcs. There are 90629 zero degree nodes ($\sim$0.1\% of the total). The average degree is $\sim$23.56. For the PLD graph we follow the same analysis procedure adopted for the host graph.

\subsection{Degree distributions}
\begin{figure}[!h]
  \begin{center}
    \includegraphics[width=3.5in]{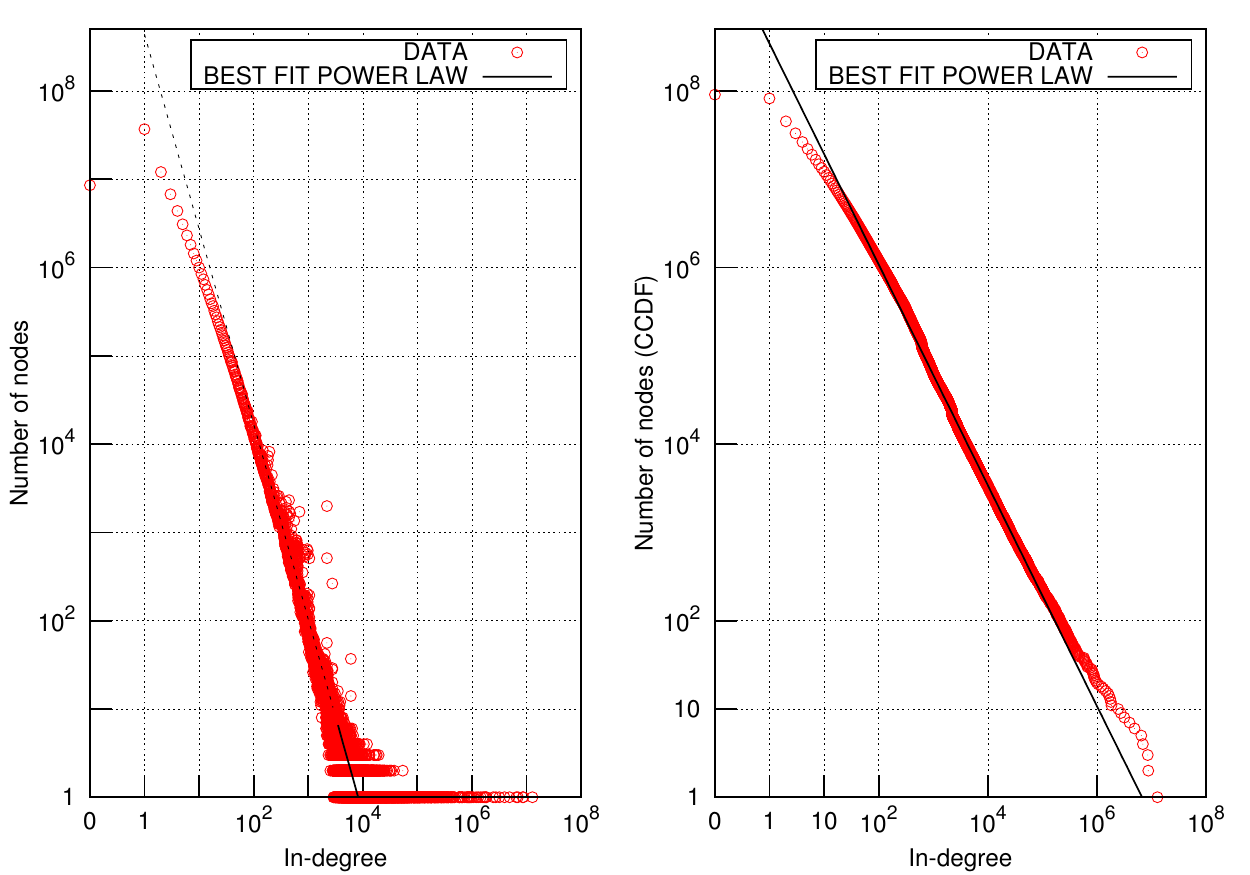}
\caption{PLD graph indegree distribution (left) and its CCDF (right). The best fit power law parameters are: $x_{min}=2858$, $\alpha=2.21\pm0.01$. The $p$-value is $0.69\pm 0.01$. }
\label{fig_in_ccd_deg_dl2017050607}
  \end{center}
\end{figure}

\begin{figure}[!h]
  \begin{center}
    \includegraphics[width=3.5in]{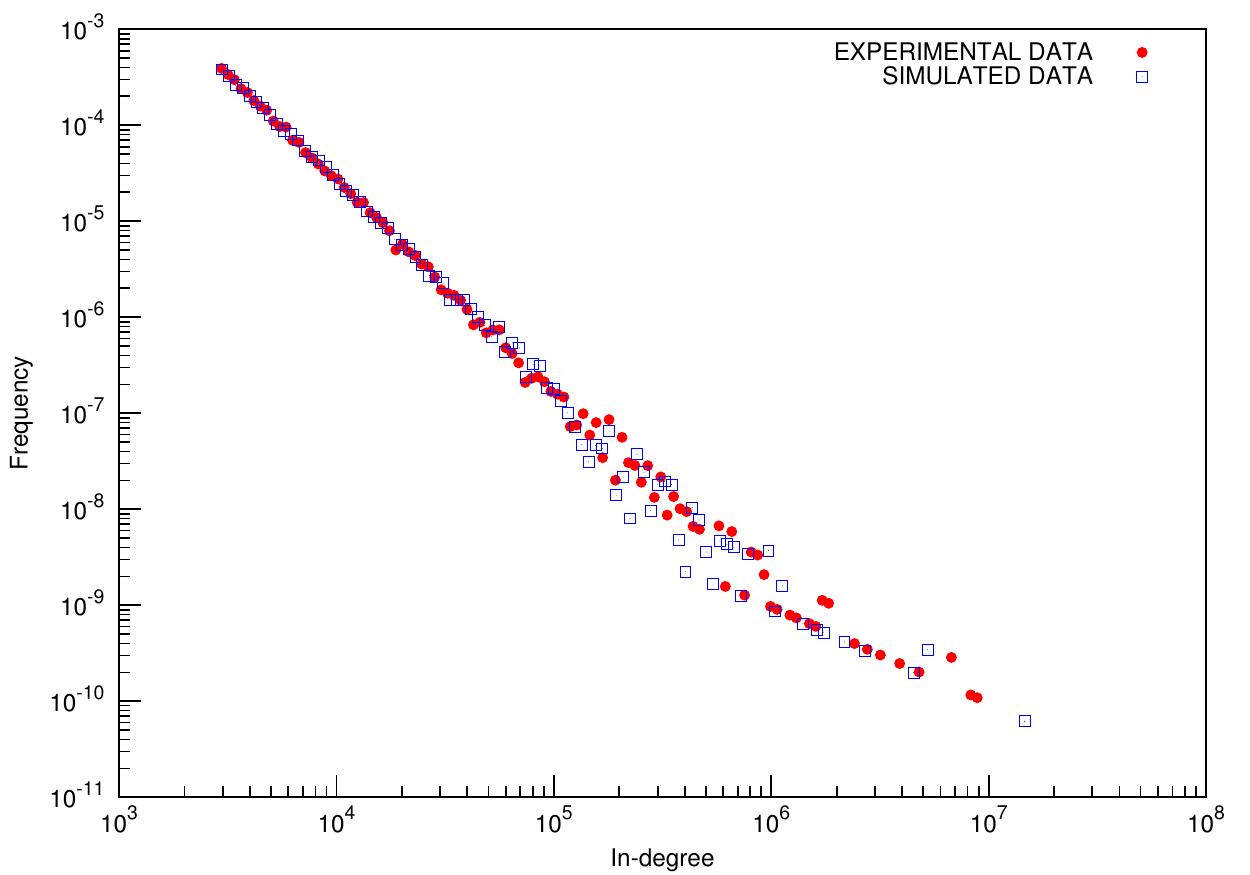}
\caption{Comparison between the indegree distribution of the PLD graph in the region $x \geq x_{min}$ and a distribution of synthetic data randomly generated from a power law having the same parameters of the detected one.  }
\label{fig_indegdl_synt_compar}
  \end{center}
\end{figure}

In Figure~\ref{fig_in_ccd_deg_dl2017050607} are shown the frequency plots of the indegree distribution and the relative CCDF. There are $\sim$8.6$\cdot 10^6$ nodes with indegree equal to zero ($\sim$9.4\% of the distribution). The maximum indegree is 12896169. There is a concavity in the region $1 \leq x \leq 30$ also visible in the CCDF plot. There are high spikes in region $2000 \leq x \leq 6000$ and even if the tail of the CCDF is not linear the best fit power law parameters are  $x_{min}=2858$, $\alpha = 2.21\pm0.01$ and $p = 069\pm0.1$. For the first time we observe a statistical evidence of a power law tail. The region $x \geq x_{min}$ contains $\sim$0.02\% of the distribution. We note that the presence of spikes is not a sufficient condition for excluding a power law as well as a linear shape of the log-log plot does not imply it. 
As a further check we compare the distribution of the experimental data in the region $x \geq x_{min}$ with that of a synthetic dataset containing the same number of samples randomly generated from  a power law  with the same parameters of the detected one. There is a good agreement between the distributions of the two datasets as shown in Figure~\ref{fig_indegdl_synt_compar}.

As in the case of the host graph analysis we now compare the power law with alternative models. The results are shown in Table~\ref{tab_compare_discrete_cont_indegdl}. Both the discrete and continuous fit calculations indicate that  none of the tested models has statistical significance as better alternative to the power law. 

\begin{table}[htbp]
\begin{center}
\resizebox{\columnwidth}{!}{ 
\begin{tabular}{c|c|c|c}
\multicolumn{4}{c}{discrete fit indegree PLD graph}\\
\hline
\hline
 \boldmath{$f$} & \boldmath{$R(pl/f)$} & \boldmath{$q$} & \makecell{\textbf{statistical plausibility of} \boldmath{$f$}\\ \textbf{as alternative to the power law}} \\
\hline
$exp$   &       9.375661 &     0 &     none \\
\hline
$logn$ &        1.558554 &    0.119102 &      undecidable \\
\hline
$tpl$ &         0.055049 &     0.988221 &      undecidable \\
\hline
$sexp$ &        76.482004 &     0 &     none \\
\hline
\hline
\boldmath{$f_A/f_B$} &  \boldmath{$R(f_A/f_B)$} & \boldmath{$q$} & \textbf{comment} \\
\hline
$logn/tpl$ &    -1.548861 &      0.121415 &      none of the tested models is favorite  \\
\end{tabular}
}
\end{center}
\begin{center}
\resizebox{\columnwidth}{!}{
\begin{tabular}{c|c|c|c}
\multicolumn{4}{c}{continuous fit indegree PLD graph}\\
\hline
\hline
 \boldmath{$f$} & \boldmath{$R(pl/f)$} & \boldmath{$q$} & \makecell{\textbf{statistical plausibility of} \boldmath{$f$}\\ \textbf{as alternative to the power law}} \\
\hline
$exp$   &       9.269584 &     0 &     none \\
\hline
$logn$ &        1.278359 &    0.201123 &      undecidable \\
\hline
$tpl$ &         0.009616 &     0.99352 &      undecidable \\
\hline
$sexp$ &        6.13434 &     0 &     none \\
\hline
\hline
\boldmath{$f_A/f_B$} &  \boldmath{$R(f_A/f_B)$} & \boldmath{$q$} & \textbf{comment} \\
\hline
$logn/tpl$ &    -1.298115 &      0.194248 &      none of the tested models is favorite  \\
\end{tabular}
}
\end{center}
\caption{Results of the likelihood ratio test for the indegree distribution of the PLD graph. Both the discrete and continuous fit calculations indicate that none of the tested models can be considered  a plausible  alternative to the power law.} 
\label{tab_compare_discrete_cont_indegdl}
\end{table}

\begin{figure}[!h]
  \begin{center}
    \includegraphics[width=3.5in]{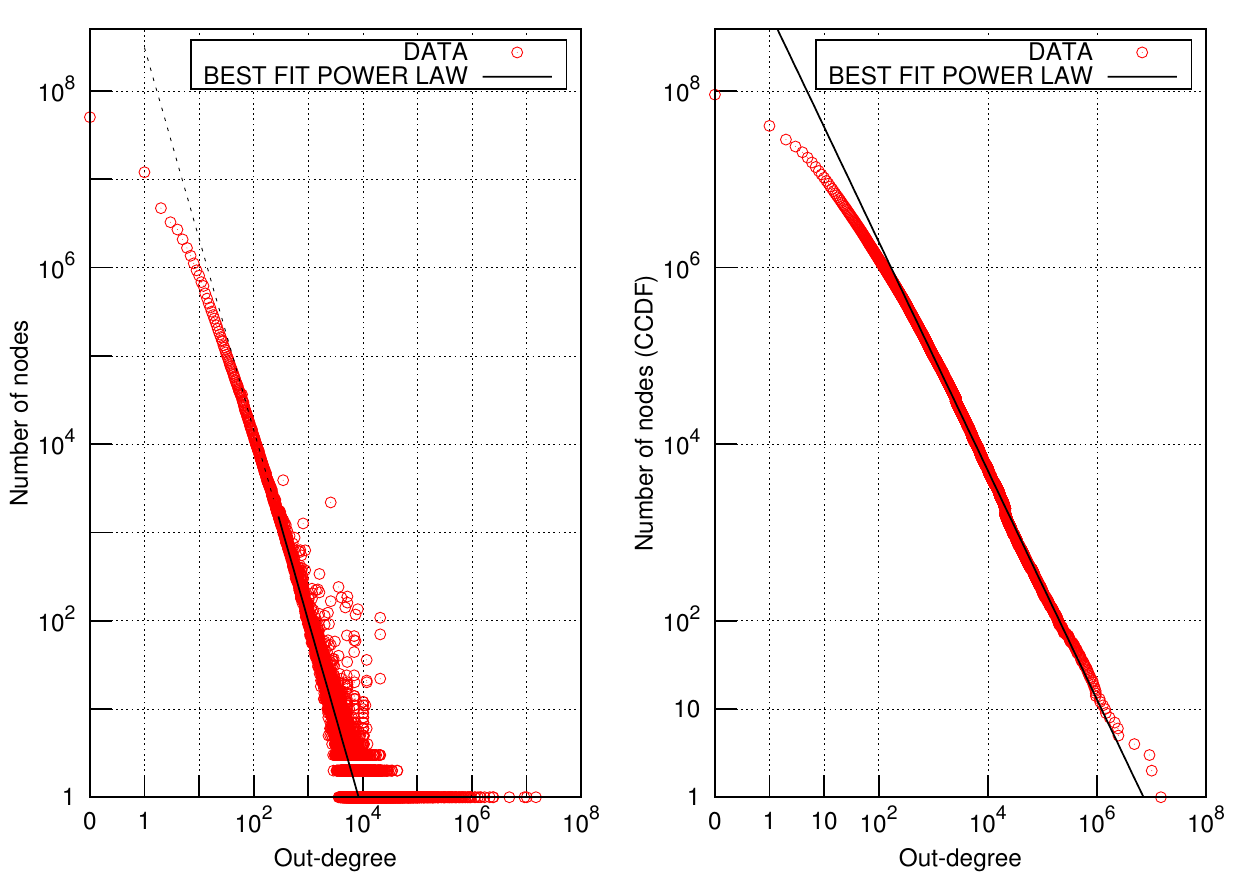}
\caption{PLD graph outdegree distribution (left) and its CCDF (right). The best fit power law parameters are: $x_{min}=279$, $\alpha=2.164\pm0.002$. The $p$-value is $0.00\pm 0.01$. }
\label{fig_out_ccd_deg_dl2017050607}
  \end{center}
\end{figure}

We now examine the outdegree distribution whose plot along with the one of its CCDF is shown in Figure~\ref{fig_out_ccd_deg_dl2017050607}. The number of nodes with outdegree equal to zero is 50659245. The most part are dangling nodes  and constitute $\sim$55.7\% of the distribution. The maximum outdegree is 14903607. There is a concavity in the region $1 \leq x \leq 60$ which is also evident in the CCDF plot. In the region $230 \leq x \leq 20000$ there are spikes. The best fit power law parameters are: $x_{min}=279$, $\alpha=2.164\pm0.002$. The $p$-value is $0.00\pm0.01$ indicating that the tail is not power law. The  points in the region $x \geq x_{min}$  are $\sim$0.5\% of the distribution.  

\begin{table}[!h]
\begin{center}
\resizebox{\columnwidth}{!}{ 
\begin{tabular}{c|c|c|c}
\multicolumn{4}{c}{discrete fit outdegree PLD graph}\\
\hline
\hline
 \boldmath{$f$} & \boldmath{$R(pl/f)$} & \boldmath{$q$} & \makecell{\textbf{statistical plausibility of} \boldmath{$f$}\\ \textbf{as alternative to the power law}} \\
\hline
$exp$   &       149.413009 &     0 &     none \\
\hline
$logn$ &        -59.471894 &    0 &      strong \\
\hline
$tpl$ &         -4.724054 &     0 &      strong \\
\hline
$sexp$ &        17.653723 &     0 &     none \\
\hline
\hline
\boldmath{$f_A/f_B$} &  \boldmath{$R(f_A/f_B)$} & \boldmath{$q$} & \textbf{comment} \\
\hline
$logn/tpl$ &    10.801170 &      0 &      strong support for the $logn$  \\
\end{tabular}
}
\end{center}
\begin{center}
\resizebox{\columnwidth}{!}{
\begin{tabular}{c|c|c|c}
\multicolumn{4}{c}{continuous fit outdegree PLD graph}\\
\hline
\hline
 \boldmath{$f$} & \boldmath{$R(pl/f)$} & \boldmath{$q$} & \makecell{\textbf{statistical plausibility of} \boldmath{$f$}\\ \textbf{as alternative to the power law}} \\
\hline
$exp$   &       14.338776 &     0 &     none \\
\hline
$logn$ &        -17.432687 &    0 &      strong \\
\hline
$tpl$ &         -1.618388 &     0 &      strong \\
\hline
$sexp$ &        2.110114 &     0.034849 &     none \\
\hline
\hline
\boldmath{$f_A/f_B$} &  \boldmath{$R(f_A/f_B)$} & \boldmath{$q$} & \textbf{comment} \\
\hline
$logn/tpl$ &    12.103655 &      0 &      strong support for the $logn$  \\
\end{tabular}
}
\end{center}
\caption{Results of the likelihood ratio test for the outdegree distribution of the PLD graph. The lognormal is the best alternative among all tested models.}
\label{tab_compare_discrete_cont_outdegdl}
\end{table}

For the outdegree distribution both the discrete and continuous fit calculations indicate strong support for the lognormal as the best alternative model to the power law, as shown in Table~\ref{tab_compare_discrete_cont_outdegdl}.

\subsection{Components}
The fraction of nodes in the largest WCC and SCC of the PLD graph are $\sim$99.4\% and $\sim$32.7\% respectively. The difference between these sizes has been reduced for the PLD graph. The largest WCC ans SCC of the PLD graph analyzed by Meusel {\itshape et al} contain $\sim$91.8\% and $\sim$51.9\% of all nodes. Also for the PLD graph this analysis confirms the presence of a giant WCC.

\begin{figure}[htbp]
  \begin{center}
    \includegraphics[width=3.5in]{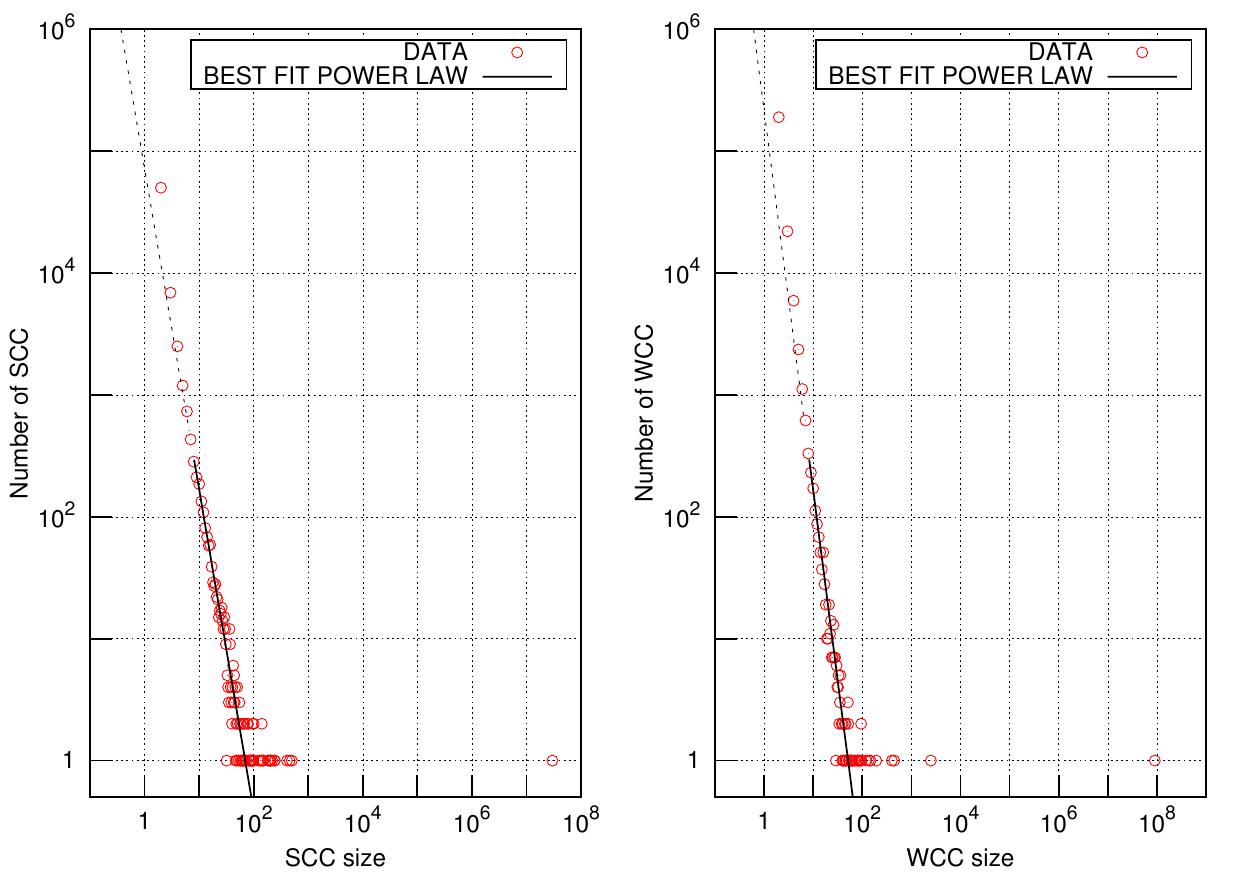}
\caption{SCC (left) and WCC (right) size distributions of the PLD graph. The best fit power law parameters for the SCC are: $x_{min}=7$, $\alpha = 2.63\pm0.04$ and $p=0.41\pm0.01$. For the WCC we have $x_{min}=8$, $\alpha = 3.12\pm0.06$ and $p=0.34\pm0.01$. Because $p > 0.1$ for both distributions the power law hypothesis has statistical support.   }
\label{fig_scc_wcc_dl2017050607}
  \end{center}
\end{figure}

In Figure~\ref{fig_scc_wcc_dl2017050607} are shown the distributions of the sizes of the SCC and WCC of the PLD graph.

The best fit power law parameter of the SCC distribution are: $x_{min}=7$, $\alpha=2.63\pm0.04$ and $p=0.41\pm0.01$. The region $x \geq x_{min}$ covers $\sim$3.2\% of the distribution.

For the WCC we obtain: $x_{min}=8$, $\alpha=3.12\pm0.06$ and $p=0.34\pm0.01$. The region $x \geq x_{min}$ covers $\sim$0.6\% of the distribution.

Both the SCC and WCC distributions have $p > 0.1$ indicating that their tails are very likely  power law. 
    
\begin{figure}[htbp]
\begin{center}
    \includegraphics[width=3.5in]{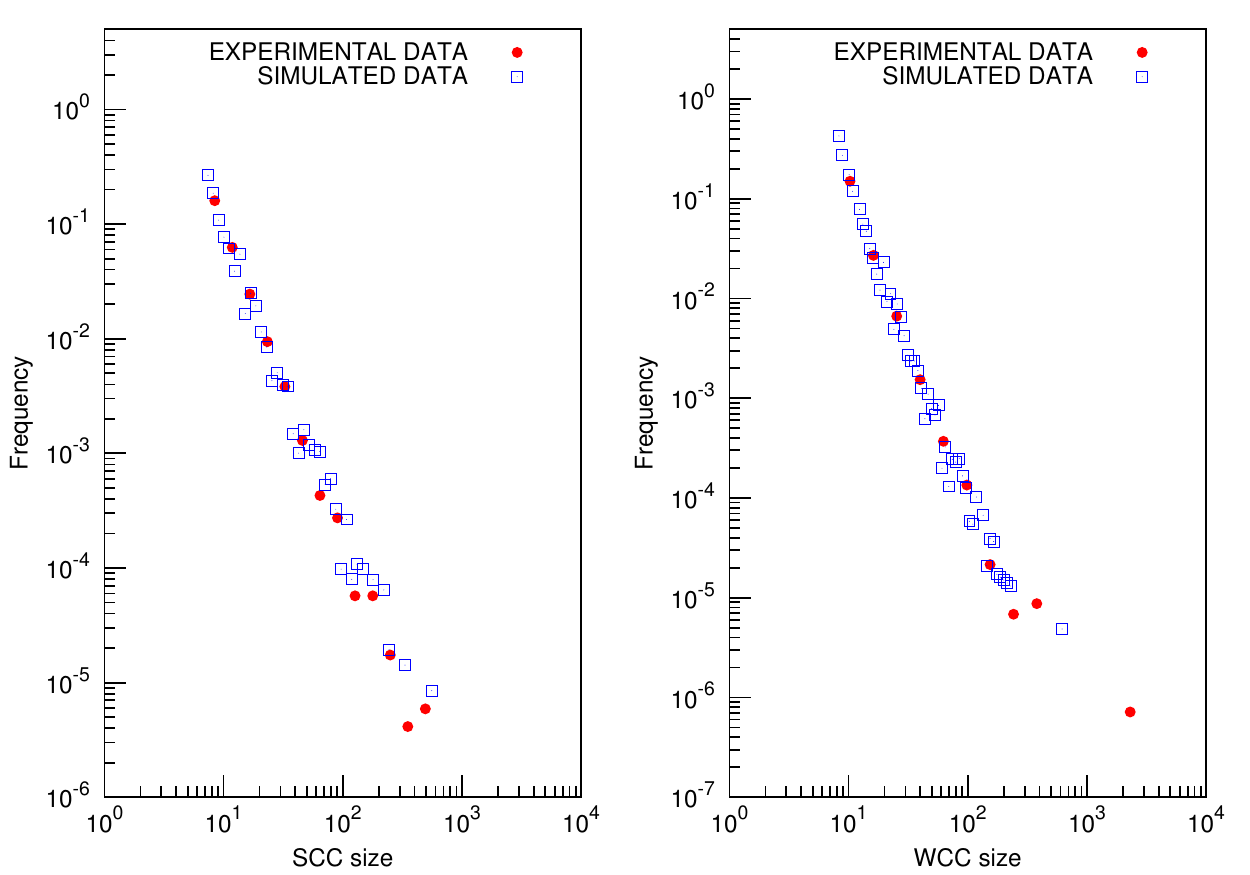}
\caption{Comparison between the SCC (left) and WCC (right) size distributions of the PLD graph in the region $x \geq x_{min}$ and that of synthetic data randomly generated from power laws with the same parameters of the detected ones.    }
\label{fig_exp_synt_scc_wcc_dl2017050607}
  \end{center}
\end{figure}

As in the case of the indegree, we make a further check of the power law hypothesis by comparing the two distributions with that of synthetic datasets in the region $x \geq x_{min}$ and find a good agreement as shown in Figure~\ref{fig_exp_synt_scc_wcc_dl2017050607}.

From Table~\ref{tab_compare_discrete_cont_sccdl} and~\ref{tab_compare_discrete_cont_wccdl} we see that the only  indication of a statistically supported model alternative to the power law comes from the  continuous fit of the SCC distribution which gives strong support to the lognormal.

\begin{table}[!h]
\begin{center}
\resizebox{\columnwidth}{!}{ 
\begin{tabular}{c|c|c|c}
\multicolumn{4}{c}{discrete fit SCC PLD graph}\\
\hline
\hline
 \boldmath{$f$} & \boldmath{$R(pl/f)$} & \boldmath{$q$} & \makecell{\textbf{statistical plausibility of} \boldmath{$f$}\\ \textbf{as alternative to the power law}} \\
\hline
$exp$   &       1.755426 &     0.079186 &     none \\
\hline
$logn$ &        0.320868 &    0.748310 &      undecidable \\
\hline
$tpl$ &         0.999271 &     0 &      none \\
\hline
$sexp$ &        46.723807 &     0 &     none \\
\hline
\end{tabular}
}
\end{center}
\begin{center}
\resizebox{\columnwidth}{!}{ 
\begin{tabular}{c|c|c|c}
\multicolumn{4}{c}{continuous fit SCC PLD graph}\\
\hline
\hline
 \boldmath{$f$} & \boldmath{$R(pl/f)$} & \boldmath{$q$} & \makecell{\textbf{statistical plausibility of} \boldmath{$f$}\\ \textbf{as alternative to the power law}} \\
\hline
$exp$   &       6.59299 &     0 &     none \\
\hline
$logn$ &        -241911.686943 &    0 &      strong \\
\hline
$tpl$ &         -241911.686659 &     0 &      strong \\
\hline
$sexp$ &        54.903007 &     0 &     none \\
\hline
\hline
\boldmath{$f_A/f_B$} &  \boldmath{$R(f_A/f_B)$} & \boldmath{$q$} & \textbf{comment} \\
\hline
$logn/tpl$ &    241911.686564 &      0 &      strong support for the $logn$  \\
\end{tabular}
}
\end{center}
\caption{Results of the likelihood ratio test for the size distribution of the SCC of the PLD graph.  Discrete fit: none of the tested  models can be considered  a statistically plausible alternative to the power law. Continuous fit: the lognormal is the most statistically plausible alternative to the power law among all tested models. }
\label{tab_compare_discrete_cont_sccdl}
\end{table}

\begin{table}[!h]
\begin{center}
\resizebox{\columnwidth}{!}{ 
\begin{tabular}{c|c|c|c}
\multicolumn{4}{c}{discrete fit WCC PLD graph}\\
\hline
\hline
 \boldmath{$f$} & \boldmath{$R(pl/f)$} & \boldmath{$q$} & \makecell{\textbf{statistical plausibility of} \boldmath{$f$}\\ \textbf{as alternative to the power law}} \\
\hline
$exp$   &       1.786023 &     0.074096 &     none \\
\hline
$logn$ &        0.431812 &    0.665878 &      undecidable \\
\hline
$tpl$ &         0.139838 &     0.530106 &      undecidable \\
\hline
$sexp$ &        45.928895 &     0 &     none \\
\hline
\hline
\boldmath{$f_A/f_B$} &  \boldmath{$R(f_A/f_B)$} & \boldmath{$q$} & \textbf{comment} \\
\hline
$logn/tpl$ &    -0.100918 &      0.919616 &      none of the tested models is favorite  \\
\end{tabular}
}
\end{center}
\begin{center}
\resizebox{\columnwidth}{!}{ 
\begin{tabular}{c|c|c|c}
\multicolumn{4}{c}{continuous fit WCC PLD graph}\\
\hline
\hline
 \boldmath{$f$} & \boldmath{$R(pl/f)$} & \boldmath{$q$} & \makecell{\textbf{statistical plausibility of} \boldmath{$f$}\\ \textbf{as alternative to the power law}} \\
\hline
$exp$   &       10.383199 &     0 &     none \\
\hline
$logn$ &        1.125393 &    0.260423 &       undecidable \\
\hline
$tpl$ &         0.999392 &     0 &       none \\
\hline
$sexp$ &        1.364604 &     0.172378 &     undecidable \\
\hline
\hline
\boldmath{$f_A/f_B$} &  \boldmath{$R(f_A/f_B)$} & \boldmath{$q$} & \textbf{comment} \\
\hline
$logn/sexp$ &    1.366401 &      0.171813 &      none of the tested models is favorite  \\
\end{tabular}
}
\end{center}
\caption{Results of the likelihood ratio test for the size distribution of the WCC of the PLD graph. Both the discrete and continuous fit calculations indicate that none of the tested  models can be considered a statistically plausible alternative to the power law. }
\label{tab_compare_discrete_cont_wccdl}
\end{table}

\subsection{Distances and diameters}
\begin{figure}[htbp]
  \begin{center}
    \includegraphics[width=3.5in]{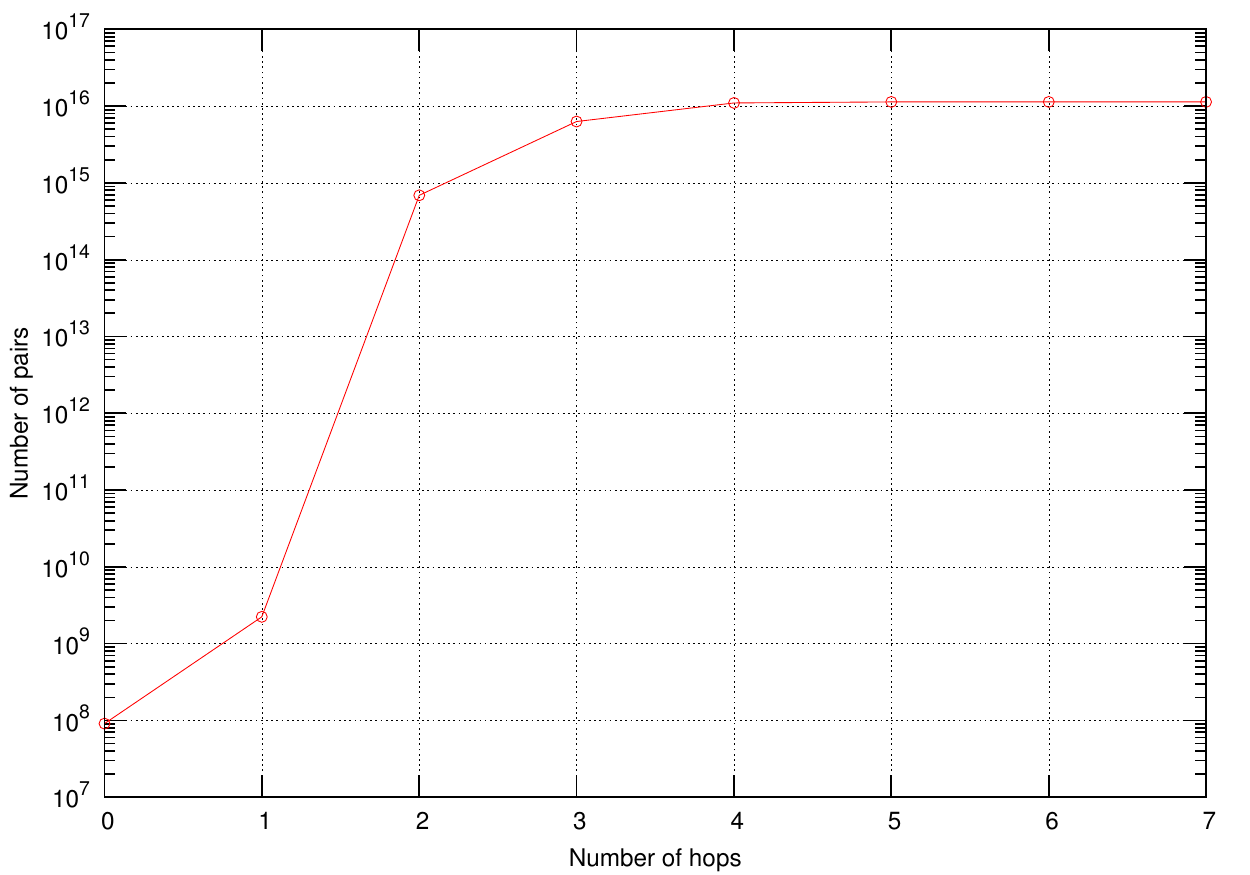}
\caption{Cumulative distribution of the shortest path lengths of the PLD graph. The Y-axis shows $N(h)$, the number of pairs of nodes with distance within $h$ hops. The effective diameter is $3.8\pm0.4$. }
\label{fig_diam_dl2017050607}
  \end{center}
\end{figure}

In the PLD graph $\sim$90\% of all pairs of nodes have distance within $3.8\pm0.4$. In Figure~\ref{fig_diam_dl2017050607} is shown the hop plot of the PLD graph. The lower bound of the full diameter estimated with a BFS using 10000 random starting nodes is 34.

\section{Conclusion}\label{sez_conclusion}
The results obtained so far are summarized in Table ~\ref{tab_summary}.
The first observation is that from this analysis we infer that there is no statistical evidence of power law tails on host level for the distributions of degree and sizes of SCC and WCC.  From the comparison between the power law and the models reported in Table~\ref{tab_models}, we find that for the host graph the lognormal is the most statistically plausible alternative. Power laws emerge on PLD aggregation for indegree, SCC and WCC size distributions. It is interesting to note that even in the case of the PLD graph the lognormal is the only model which is not ruled out by the likelihood ratio test. Of course, there might be other models which fit better the data.

\begin{table}[!h]
\begin{center}
\resizebox{\columnwidth}{!}{
\begin{tabular}{c|c|c}
May-June-July 2017  & \textbf{Host graph}   & \textbf{PLD graph}    \\
\hline
\hline
\# nodes   &       1306661614 &     91034128  \\
\hline
\# arcs &        5268397861 &    1071173924   \\
\hline
indegree & &  \\
$\alpha$ & $2.193\pm0.001$ & $2.21\pm0.01$ \\
$x_{min}$ & 250  & 2858 \\
largest & 23055296 & 12896169 \\
\hline
outdegree & &  \\
$\alpha$ & $2.3242\pm0.0001$ & $2.164\pm0.002$ \\
$x_{min}$ & 23  & 279 \\
largest & 15090917 & 14903607 \\
\hline
SCC & &  \\
$\alpha$ & $2.367\pm0.005$ & $2.63\pm0.04$ \\
$x_{min}$ & 4 & 7 \\
largest & $\sim$4.5\% & $\sim$32.7\% \\
\hline
WCC & &  \\
$\alpha$ & $1.684\pm0.001$ & $3.12\pm0.06$ \\
$x_{min}$ & 22 & 8 \\
largest & $\sim$99.7\% & $\sim$99.4\% \\
\hline
Effective diameter & $5.6\pm0.6$ & $3.8\pm0.4$\\
\hline
Full diameter (lower bound) & 970 & 34\\
\hline
\end{tabular}
}
\end{center}
\begin{center}
\resizebox{\columnwidth}{!}{
\begin{tabular}{c|c|c|c}
\multicolumn{4}{c}{Host graph}\\
\hline
\hline
 \textbf{distribution} & \makecell{\textbf{statistical support} \\ \textbf{for the power law}}  & \makecell{\textbf{statistical support} \\ \textbf{for alternative models}\\ \textbf{(discrete fit)}}  & \makecell{\textbf{statistical support} \\ \textbf{for alternative models} \\ \textbf{(continuous fit)}} \\
\hline
indegree   &       none &     lognornal (strong) &     none \\
\hline
outdegree &        none &    lognornal (strong) &      lognornal (strong) \\
\hline
SCC &         none &     lognornal (strong) &      none \\
\hline
WCC &        none &     lognornal (strong) &     none \\
\hline
\end{tabular}
}
\end{center}
\begin{center}
\resizebox{\columnwidth}{!}{
\begin{tabular}{c|c|c|c}
\multicolumn{4}{c}{PLD graph}\\
\hline
\hline
\textbf{distribution} & \makecell{\textbf{statistical support} \\ \textbf{for the power law}}  & \makecell{\textbf{statistical support} \\ \textbf{for alternative models} \\ \textbf{(discrete fit)}}  & \makecell{\textbf{statistical support} \\ \textbf{for alternative models} \\ \textbf{(continuous fit)}} \\
\hline
indegree   &       yes &     none &     none \\
\hline
outdegree &        none &    lognornal (strong) &      lognornal (strong) \\
\hline
SCC &         yes &     none &      lognornal (strong) \\
\hline
WCC &        yes &     none &     none \\
\hline
\end{tabular}
}
\end{center}
\caption{Summary of the analysis of the web graph aggregated by host and PLD. }
\label{tab_summary}
\end{table}

In the analysis of Meusel {\itshape et al}. it was found no statistical evidence of power laws on page and host levels but only for the indegree distribution of the PLD graph.  Therefore the scale-free nature of the web, namely the coexistence of nodes with very low (or zero) degree and nodes with millions of links, is not necessarily a consequence of mechanisms which predict a power law form of the degree distributions.

Another observation is that while the fraction of nodes in the largest  WCC is $\sim$99\% both in the host and PLD graphs, the fraction of nodes in the largest SCC varies considerably and is $\sim$4\% in the host graph and $\sim$33\% in the PLD graph. In the analysis of Broder {\itshape et al}. the size of the largest SCC is $\sim$28\% of the whole page graph and in the  work of Meusel {\itshape et al}. it is roughly $\sim$50\%  of the whole graph for all levels of aggregation. It is possible that the size of the largest SCC is significantly affected by the crawling strategy and the processes of graph extraction and level aggregation. 

The distance between two randomly chosen nodes in the host and PLD graphs is very likely within $\sim$5.6 and $\sim$3.8 hops, respectively. This numbers should be compared with the average distance measured by Meusel {\itshape et al}. which is $\sim$5.3 in the host graph and $\sim$4.3 in the PLD graph. This small-world property of the web is particularly evident in the case of the host graph where the effective diameter remains almost the same even if the size changes by one order of magnitude.

\section{Acknowledgements}

The computing resources and the related technical support used for this work have been provided by CRESCO/ENEAGRID High Performance Computing infrastructure and its staff~\cite{eneagrid}. CRESCO/ENEAGRID High Performance Computing infrastructure is funded by ENEA, the Italian National Agency for New Technologies, Energy and Sustainable Economic Development and by Italian and European research programmes, see https://www.eneagrid.enea.it for information.

\bibliographystyle{unsrt}
\bibliography{webstructurerefs}
\end{document}